# On the hydrogen wave function in Momentum-space, Clifford algebra and the Generating function of Gegenbauer polynomial

**M. Hage-Hassan**
Université Libanaise, Faculté des Sciences Section (1)
Hadath-Beyrouth

**Abstract**

Using the quadratic transformation $R^4 \to R^3$ and the generating function method we perform the Fourier transformation of the wave function of coordinates of hydrogen atom and we find the analytic expression of the wave function in momentum space. We derive the matrix elements between the basis to 4-dimensions and integral representation of the generating functions of Gegenbauer polynomials. We find a relationship between a class of Clifford algebra and the generating functions of these polynomials.

**1-Introduction**

The problem of the hydrogen atom has played a central role in the development of quantum mechanics. Schrödinger solved his equation and found the wave function of the coordinate representation. The problem in momentum space has been reformulated by Fock [1] which led to an integral form of the Schrödinger equation. This equation is solved by projecting the three-dimensional momentum space onto the surface of a four-dimensional sphere and the eigenfunctions are then expanded in terms of spherical harmonics.

Despite the importance of Fock's work and the interest of many authors [2-8] to study the wave function in momentum space it must not hide that the direct calculation of Fourier transform of the wave function of coordinates is up till now undone and our aim in this work is to fill this gap.

The wave function of coordinates [9] is in the form $\psi_{nlm}(\vec{r}) = R_{nl}(\omega r) Y_{lm}(\theta\varphi), \omega = 2/n$. Where $R_{nl}(\omega r)$ is the radial part, $Y_{lm}(\Omega)$ is the spherical harmonic and $\Omega$ the solid angle. The difficulty for the determination of the wave function in momentum space comes from $\omega$ and the appearance of the term "r" in the exponential of the radial part. We propose to circumvent these problems by using the quadratic transformation $R^4 \to R^3$ and the generating function method where $\omega = 2/n_0$ is a constant for all the elements of the basis. After calculation we found in the expansion of the obtained function the Fourier transform of the wave functions in position space with $n = n_0$, and then we obtain the analytic expression of the wave function of hydrogen atom in momentum representation.

To not cumbersome the text by the applications we limit ourselves for the passage formula between the spherical function and Wigner's D-matrix elements of SU(2). We find an Integral representation of the generating functions of Gegenbauer polynomials and as far as I know this is a new formula.

We find also that the Bargmann's integral of a class of quadratic forms related to Clifford algebra gives as solution the generating function of Gegenbauer Polynomials.

This paper is organized as follows. In part 2 we construct the generating function for the basis of the hydrogen atom. The next section is devoted to the presentation of the connection of $R^3$ hydrogen atom and $R^4$ harmonic oscillator. In section 4 we derive the wave functions of hydrogen atom in momentum space. In section 5 we derive the passage formula between the basis of $R^4$ and the integral representation of the generating functions of Gegenbauer polynomials. In the last section we present the relation between the Clifford algebra and generating function of these polynomials.

## 2. Generating function for the basis of the hydrogen atom

The wave function of hydrogen atom in momentum representation [9] is

$$\psi_{nlm}(\vec{p}) = \frac{1}{(2\pi)^{3/2}} \int e^{-i\vec{p}\cdot\vec{r}} \psi_{nlm}(\vec{r}) d\vec{r} \tag{2.1}$$

With $R_{nl}(x)$ is the radial part
$$R_{nl}(x) = \frac{N_{nl}}{(n+l)!} x^l e^{-\frac{x}{2}} L_{n-l-1}^{2l+1}(x) \tag{2.2}$$

And
$$x = \omega r, \quad N_{nl} = \frac{2}{n^2}\sqrt{\frac{(n-l-1)!}{[(n+l)!]}}, \quad \omega = 2\delta, \delta = \frac{1}{n} \tag{2.3}$$

Where $L_\alpha^\beta(x)$ is the associated Laguerre polynomial. Atomic unit are used through the text.

**2.1 The generating function of Laguerre polynomial $L_{n-l-1}^{2l+1}(r)$**

The generating function of Laguerre polynomial is:

$$\sum_{n=0}^{\infty} \frac{z^n}{(n+r)!} L_n^{(r)}(r) = \frac{1}{(1-z)^{r+1}} e^{-\frac{z}{1-z}r}$$

From the property
$$\frac{d}{dr} L_n^{(\alpha)}(r) = -L_{n-1}^{(\alpha+1)}(r)$$

We deduce that

$$\sum_{n=0}^{\infty} \frac{z^n}{(n+r)!} L_{n-l-1}^{2l+1}(r) = \frac{(z)^{l+1}}{(1-z)^{2l+2}} \exp\left(-\frac{zr}{(1-z)}\right) \tag{2.4}$$

**2.2 The generating function of spherical harmonics**

The generating function of spherical harmonics is:

$$\frac{(\vec{a}\cdot\vec{r})^l}{2^l l!} = \left[\frac{4\pi}{2l+1}\right]^{\frac{1}{2}} \sum_m \varphi_{lm}(\xi) Y_{lm}(\vec{r}) \tag{2.5}$$

With $\vec{a}$ is a vector of length zero, $\vec{a}\cdot\vec{a} = \vec{a}^2 = 0$ and its components
$$a_1 = -\xi^2 + \eta^2, \quad a_2 = -i(\xi^2 + \eta^2), \quad a_3 = 2\xi\eta$$

With
$$\varphi_{lm}(z) = \frac{\xi^{l+m}\eta^{l-m}}{\sqrt{(l+m)!(l-m)!}}, z = (\xi,\eta)$$

### 2.3 Generating function for the basis of the hydrogen atom

We multiply $\Psi_{nlm}(\vec{r})$ by $\left(\frac{4\pi}{2l+1}\right)^{\frac{1}{2}} \frac{z^n}{N_{nl}} \alpha\varphi_{lm}(\xi)$, and summing with respect to $n,l,m$

$$G(z,\alpha\xi,\vec{r}) = \sum_{nlm}\left(\frac{4\pi}{2l+1}\right)^{\frac{1}{2}} \frac{z^n}{N_{nl}} \alpha\varphi_{lm}(\xi)\Psi_{nlm}(\vec{r}) =$$

$$\sum_{nl} e^{-(\omega r)/2} \frac{z^n}{(n+l+1)!} L_{n-l-1}^{2l+1}(\omega r) \sum_m \alpha\varphi_{lm}(\xi) Y_{lm}(\omega\vec{r})$$

Substituting (1.4) and (1.5) in the above expression we obtain:

$$G(z,\alpha\xi,\vec{r}) = \frac{z}{(1-z)^2} \exp[-\frac{\omega r(1+z)}{2(1-z)} + \alpha\frac{\omega z(\vec{a}.\vec{r})}{2(1-z)^2}] \tag{2.6}$$

## 3. The connection of $R^3$ hydrogen atom and $R^4$ harmonic oscillator

We will derive the quadratic transformation by a simple way then we determine the volume element. A summary of the connection between the wave function of hydrogen atom and harmonic oscillator is given in the following.

### 3.1 The quadratic transformation $R^4 \to R^3$

The quadratic transformation $R^4 \to R^3$ has been used first by Kustaanheimo-Steifel [10] in celestial mechanics and was used also by many authors [11-12] for the connection of $R^3$ hydrogen atom and $R^4$ harmonic oscillator. We shall derive this transformation by a simple method knowing that its derivation can be done by several ways [12-16].
Consider the relationship between the well-known Wigner's D matrix and spherical harmonics polynomials [17]

$$\sqrt{\frac{4}{2l+1}} Y_{lm}^*(\vec{r}) = D_{(0,m)}^l(z_1,\bar{z}_1,z_2,\bar{z}_2) \tag{3.1}$$

$$z_1 = u_1 + iu_2, \quad z_2 = u_3 + iu_4$$
$$r = \vec{u}^2 = u_1^2 + u_2^2 + u_3^2 + u_4^2$$

We write in terms of Euler's angles or Cayley-Klein parameterization.

$$z_1 = u_1 + iu_2 = \sqrt{r}\cos\frac{\theta}{2} e^{-\frac{i(\phi+\psi)}{2}}, \quad z_2 = u_3 + iu_4 = \sqrt{r}\sin\frac{\theta}{2} e^{-\frac{i(\phi-\psi)}{2}} \tag{3.2}$$

and $D_{(m',m)}^j(z_1,\bar{z}_1,z_2,\bar{z}_2) = u^{2j} D_{(m',m)}^j(\psi\theta\varphi), j = 1, 1/2,...$
It is important to emphasize that the elements of the matrix D are solution of Laplacian $\Delta_4$ with the parameterization of Cayley-Klein.

If we put $l = 1$ in (3, 1) we obtain the quadratic transformation $R^4 \to R^3$:
$$x = 2(u_1 u_3 + u_2 u_4) = z_1\bar{z}_2 + z_2\bar{z}_1, \quad y = 2(u_1 u_4 - u_2 u_3) = i(z_1\bar{z}_2 - z_2\bar{z}_1)$$

and $$z = u_1^2 + u_2^2 - u_3^2 - u_4^2 = z_1\bar{z}_1 - z_2\bar{z}_2 \qquad (3.3)$$

**3.2 The volume element**

We consider the transformation $(u_1, u_2, u_3, u_4) \to (r, \theta, \varphi, \psi)$

With $\quad 0 \leq \theta \leq \pi,\ 0 \leq \psi, \varphi \leq 2\pi,\ 0 \leq r \leq \infty,\ -\infty \leq u_i \leq +\infty, i = 1,\ldots,4.$

and $$d^4\vec{u} = |J| dr d\theta d\varphi d\psi$$

The calculation of the Jacobian gives $|J| = (u^2/8)\sin\theta$ but $d^3\vec{r} = r^2 dr d\theta d\varphi d\psi$

Therefore $\quad 8u^2 d\vec{u} = d\vec{r} d\psi$

And $\quad \int f(x,y,z) d^3\vec{r} = \dfrac{1}{2\pi}\int f(x,y,z) d^3\vec{r} d\psi = \dfrac{4}{\pi}\int f(x(u), y(u), z(u)) d^4\vec{u} \quad (3.4)$

**3.3 The connection of hydrogen atom and harmonic oscillator**

A quick calculation shows that the equation of the hydrogen atom

$$[-\frac{\hbar^2}{2\mu}\Delta - \frac{Ze^2}{r}]\Psi = E\Psi. \quad (\mu \text{ is the reduced mass}).$$

That may be written on the basis of harmonic oscillator in the form

$$[-\frac{\hbar^2}{2\mu}\sum_{i=1}^{4}\frac{\partial^2}{\partial u_i^2} - 4E\sum_{i=1}^{4}u_i^2]\Psi = 4Ze^2\Psi \qquad (3.5)$$

With a constraint on the eigenfunctions: $\quad \dfrac{\partial}{\partial \psi}\Psi = 0. \qquad (3.6)$

and $\quad \omega = \sqrt{-8E/\mu},\quad 4Ze^2 = \hbar\omega(n+2)$

The energy is given by: $\quad E = -2\mu\left(\dfrac{Ze^2}{\hbar(n+2)}\right)^2 \qquad (3.7)$

## 4- The wave functions of hydrogen atom in momentum space

We write first the Fourier transform in the representation (u) and with the help of Bargmann integral we determine the generating function in momentum representation. Finally the development of this function gives us the wave functions of hydrogen atom in momentum space.

**4.1 The generating function in {u} representation**

We denote the generating function by $G(z, \alpha\xi, \vec{p})$ in the representation $\{u\}$. But to determine the generating function (2.3) we must multiply by $4/\pi$ to reflect the change in the measure of integration. We write

$$\psi_{nlm}(\vec{p}) = \dfrac{1}{(2\pi)^{3/2}}\int e^{-i\vec{p}\cdot\vec{r}}\psi_{nlm}(\vec{r}) d^3\vec{r} \qquad (4.1)$$

To calculate this expression we must write (4.1) in the (u) representation using the formula (3.4):

$$\psi_{nlm}(\vec{p}) = \frac{4}{\pi} \frac{1}{(2\pi)^{3/2}} \int e^{-i\vec{p}.\vec{r}} \psi_{nlm}(\vec{r}) u^2 d^4\vec{u} \qquad (4.2)$$

In the expression $\psi_{nlm}(\vec{p})$ there is the term $u^2$ for that we consider a new generating function:

$$G(z,\alpha\xi,\vec{p},\beta) = \frac{1}{(2\pi)^{3/2}} \frac{4}{\pi} \frac{z}{(1-z)^2} \times$$

$$\int e^{-i\vec{p}.\vec{r}} \exp[-\frac{\omega r(1+z)}{2(1-z)} + \frac{\alpha\omega z(\vec{a}.\vec{r})}{2(1-z)^2}] e^{-\beta u^2} d^4\vec{u} \qquad (4.3)$$

We assume that $\beta \geq 0$ therefore there is no problem of convergence.

We write then $\quad [-\frac{\partial}{\partial \beta} G(z,\alpha\xi,\vec{p},\beta)]\big|_{\beta=0} = G(z,\alpha\xi,\vec{p}) \qquad (4.4)$

With $\quad G(z,\alpha\xi,\vec{p}) = \sum_{nlm} \left(\frac{4\pi}{2l+1}\right)^{\frac{1}{2}} \frac{z^n}{N_{nl}} \alpha^l \varphi_{lm}(\xi) \psi_{nlm}(p) \qquad (4.5)$

**4.2 The generating function of momentum-space**

We can do the integration of (4.3) by a direct calculation with the variables (u) or more quickly using the Bargmann integral [18]

$$(1/\pi^n) \int \prod_{i=1}^{n} d^2 v_i \exp(-\bar{v}^t X v + A^t v + \bar{v}^t B) = (\det X)^{-1} \exp(A^t X^{-1} B) \qquad (4.6)$$

With $v = (v_1, v_2, .., v_n)$

We have $-i\vec{p}.\vec{r} = -ip_x(z_1\bar{z}_2 + z_2\bar{z}_1) + p_y(z_1\bar{z}_2 - z_2\bar{z}_1) - ip_z(z_1\bar{z}_1 - z_2\bar{z}_2)$

$$\vec{a}.\vec{r} = a_x(z_1\bar{z}_2 + z_2\bar{z}_1) + ia_y(z_1\bar{z}_2 - z_2\bar{z}_1) + a_z(z_1\bar{z}_1 - z_2\bar{z}_2) \qquad (4.7)$$

$$r = z_1\bar{z}_1 + z_2\bar{z}_2$$

We obtain then

$$X = \begin{pmatrix} \frac{\omega(1+z)}{2(1-z)} + \beta - ip_z + \frac{\omega z}{2(1-z)^2} a_z & -ip_x + \frac{\omega z}{2(1-z)^2} a_x - p_y + \frac{1}{i}\frac{\omega z}{2(1-z)^2} a_y \\ -ip_x + \frac{\omega z}{2(1-z)^2} a_x + p_y - \frac{1}{i}\frac{\omega z}{2(1-z)^2} a_y & \frac{\omega(1+z)}{(1-z)} + \beta + ip_z - \frac{\omega z}{2(1-z)^2} a_z \end{pmatrix}$$

Because $\vec{a}^2 = 0$ we deduce that:

$$\det(X) = [\left(\frac{\delta(1+z)}{(1-z)} + \beta\right)^2 + \vec{p}^2 + i\alpha \frac{2\delta z}{(1-z)^2} \vec{a}.\vec{p}], \quad \delta = \frac{\omega}{2} \qquad (4.8)$$

We find therefore the generating function

$$G(z,\alpha\xi,\vec{p},\beta) = \frac{2}{\sqrt{2\pi}} \frac{z}{[(\delta(1+z)+\beta(1-z))^2 + (1-z)^2 \vec{p}^2 + 2\alpha i \delta z \vec{a}.\vec{p}]} \qquad (4.9)$$

In applying the relation (4.4) we find the generating function $G(z,\alpha\xi,\vec{p})$

$$G(z,\alpha\xi,\vec{p}) = \frac{4\delta}{\sqrt{2\pi}} \frac{z(1-z^2)}{[(\delta(1+z))^2 + (1-z)^2 \vec{p}^2 + 2\alpha i \delta z \vec{a}.\vec{p}]^2} \qquad (4.10)$$

### 4.3 The wave functions in momentum-space

We drive the basis of momentum-space using the formula

$$[\varphi_{jm}(\partial/\partial\xi)\frac{1}{n!}\frac{\partial^n}{\partial z^n}\frac{1}{l!}\frac{\partial^l}{\partial\alpha^l}G(z,\alpha\xi,\vec{p})]_0 = \left(\frac{4\pi}{2l+1}\right)^{\frac{1}{2}}\frac{1}{N_{nl}}\psi_{nlm}(\vec{p}) \quad (4.11)$$

In this case we must take $\delta=1/n$ and to execute the calculations we proceed by step:

1 - <u>Derivation with respect to $\alpha$</u>

$$[\frac{1}{l!}\frac{\partial^l}{\partial\alpha^l}G(z,\alpha\xi,\vec{p})]_0 = (i)^l\frac{(l+1)!}{\sqrt{2\pi}}\times(4\delta)^{l+1}$$

$$\frac{(1-z^2)z^{l+1}}{[(\delta(1+z))^2+(1-z)^2\vec{p}^2]^{l+2}}\frac{(\vec{a}.\vec{p})^l}{2^l l!} \quad (4.12)$$

We have $(\delta(1+z))^2+(1-z)^2\vec{p}^2 = ((\vec{p}^2+\delta^2)-2z(\vec{p}^2-\delta^2)+z^2(\vec{p}^2+\delta^2))$

$$= (\vec{p}^2+\delta^2)[1-2zx+z^2], \quad x=\left(\frac{\vec{p}^2-\delta^2}{\vec{p}^2+\delta^2}\right)$$

We deduce that

$$[\frac{1}{l!}\frac{\partial^l}{\partial\alpha^l}G(z,\alpha\xi,\vec{p})]_0 = (i)^l\frac{(l+1)!}{\sqrt{2\pi}}\times\frac{(4\delta)^{l+1}}{(\vec{p}^2+\delta^2)^{l+2}}[\frac{(1-z^2)z^{l+1}}{[1-2zx+z^2]^{l+2}}]\frac{(\vec{a}.\vec{p})^l}{2^l l!} \quad (4.13)$$

2- <u>Derivation with respect to $z$</u>

Using the familiar formula for the generating function of Gegenbauer polynomials

$$(1-2rt+r^2)^{-\alpha} = \sum_{m=0}^{\infty}r^m C_m^{\alpha}(t) \quad (4.14)$$

We write $\quad\dfrac{(1-z^2)z^{l+1}}{[1-2zx+z^2]^{l+2}} = \sum_{m=0}^{\infty}(1-z^2)z^{l+1}C_m^{l+2}(x)$

$$= \sum_{n=1}^{\infty}z^n\left[C_{n-l-1}^{l+2}(x)-C_{n-l-3}^{l+2}(x)\right]$$

With $m+l+1=n$, $m+l+3=n$ and $\delta=1/n$ therefore

$$[\frac{1}{n!}\frac{\partial^n}{\partial z^n}\frac{1}{l!}\frac{\partial^l}{\partial\alpha^l}G(z,\alpha\xi,\vec{p})]_0 = (i)^l\frac{(l+1)!}{\sqrt{2\pi}}\times\frac{(4\delta)^{l+1}}{(\vec{p}^2+\delta^2)^{l+2}}\times$$

$$\left[C_{n-l-1}^{l+2}(x)-C_{n-l-3}^{l+2}(x)\right]\frac{(\vec{a}.\vec{p})^l}{2^l l!} \quad (4.15)$$

Put $\quad \vec{y}=(y_1,y_2,y_3,y_4)=(\dfrac{\delta p_x}{(\vec{p}^2+\delta^2)},\dfrac{\delta p_x}{(\vec{p}^2+\delta^2)},\dfrac{\delta p_x}{(\vec{p}^2+\delta^2)},\dfrac{(\vec{p}^2-\delta^2)}{(\vec{p}^2+\delta^2)})$

We obtain $\quad \vec{y}.\vec{y}=1$

Thus we find the transformation introduced by Fock.

3- <u>Derivation with respect to $\varphi_{jm}(\partial/\partial\xi)$</u>

By using the formula (2.5) we get the following expression

$$[\varphi_{jm}(\partial/\partial\xi)\frac{1}{n!}\frac{\partial^n}{\partial z^n}\frac{1}{l!}\frac{\partial^l}{\partial \alpha^l}G(z,\alpha\xi,\vec{p})]\Big|_0 = (i)^l \frac{(l+1)!}{\sqrt{2\pi}} \times \frac{(4\delta)^{l+1}}{(\vec{p}^2+\delta^2)^{l+2}} \times$$
$$[C_{n-l-1}^{l+2}(x) - C_{n-l-3}^{l+2}(x)]Y_{lm}(\vec{p}) \qquad (4.16)$$

4- <u>The wave functions in momentum space</u>

The comparisons of (4.16) and (4.12) give us the result:

$$\psi_{nlm}(\vec{p}) = (i)^l N_{nl} a^{-3/2} \frac{(l+1)!}{\sqrt{2\pi}} \times (4\delta)^{l+1} \frac{[C_{n-l-1}^{l+2}(x) - C_{n-l-3}^{l+2}(x)]}{(\vec{p}^2+\delta^2)^{l+2}} Y_{lm}(\vec{p}) \qquad (4.17)$$

And with the help of the recurrences formula [21]

$$(n+\alpha)C_{n+1}^{(\alpha-1)}(x) = (\alpha-1)[C_{n+1}^{(\alpha)}(x) - C_{n-1}^{(\alpha)}]$$

We derive finally the wave functions in momentum space:

$$\psi_{nlm}(\vec{p}) = (i)^l N_{nl} \frac{(l)!}{\sqrt{2\pi}} \times \frac{n(4\delta)^{l+1}}{(\vec{p}^2+\delta^2)^{l+2}} C_{n-l-1}^{l+1}\left(\frac{\vec{p}^2-\delta^2}{\vec{p}^2+\delta^2}\right) Y_{lm}(\vec{p}) \qquad (4.18)$$

It is clear that we obtain by an elementary method not only the wave function in momentum representation [4] but also the phase factor.

## 5- Passage formulas between the basis of $R^4$ and between the generating functions of Gegenbauer.

In order do not cumbersome the text with the applications I am restricted only to the calculation of the passage matrix elements from the spherical basis to the basis which elements are $D_{(m',m)}^j(\psi\theta\varphi)$ and the mapping between the generating functions of Gegenbauer polynomials.

If we consider the spherical parameterization

$$x = v\sin\chi\sin\theta\cos\varphi, \, y = v\sin\chi\sin\theta\sin\varphi, \, z = v\sin\chi\cos\theta, \, q = v\cos\chi \qquad (5.1)$$

The Laplacian $\Delta_4$ have solutions the spherical functions [2-19]

$$Y_{nlm}(\vec{v}) = 2^{l+1} v^{n-l+1} l! \left(\frac{n(n-l-1)!}{2\pi(n+l)!}\right)^{\frac{1}{2}} C_{n-l-1}^{l+1}(\cos\chi) Y_{lm}(\vec{r}), \qquad (5.2)$$

With $\qquad \vec{v} = (\vec{r},q), \vec{r} = (x,y,z)$

**5.2 Passage formulas between the bases of $R^4$**

We consider the generating function [20] of Wigner's D-matrix:

$$\exp[(z'_1 \quad z'_2)\begin{pmatrix} q+iz & x+iy \\ -x+iy & q-iy \end{pmatrix}\begin{pmatrix} z_1 \\ z_2 \end{pmatrix}] = \sum_{jmm'} \varphi_{jm'}(z')\varphi_{jm}(z)D_{(m',m)}^j(rU_2),$$
$$\vec{v}^2 = q^2 + x^2 + y^2 + z^2. \qquad (5.3)$$

By replacing in the formula $(z_1 \quad z_2)$ by $(\bar{z}'_1 \quad \bar{z}'_2)$ we have

$$\begin{pmatrix} z_1' & z_2' \end{pmatrix} \begin{pmatrix} q+iz & x+iy \\ -x+iy & q-iz \end{pmatrix} \begin{pmatrix} \bar{z}_1' \\ \bar{z}_2' \end{pmatrix} = qq'+ixx'+iyy'+izz' = qq'+i\vec{r}.\vec{r}' \qquad (5.4)$$

We find then the quadratic transformation $R^4 \to R^3$, $(u'_1, u'_2, u'_3, u'_4) \to (x', y', z')$

With $\quad \vec{r} = (x, y, z)$, $\vec{r}' = (x', y', z')$, $q' = r' = u_1'^2 + u_2'^2 + u_3'^2 + u_4'^2$

$$\exp[qq'+i\vec{r}.\vec{r}'] = \sum_{jmm'} \varphi_{jm'}(\bar{z}')\varphi_{jm}(z')r^{2j} D^j_{(m',m)}(U_2) \qquad (5.5)$$

A second form of (5.5) can be done through the development of the wave free, then the generating function of Gegenbauer polynomials and Legendre duplication formula.

We develop $e^{i\vec{r}.\vec{r}}$ on the spherical harmonics basis

$$e^{i\vec{r}'.\vec{r}} = 4\pi \sum_{l=0}^{\infty} \sum_{m=-l}^{+l} i^l j_l(rr') Y^*_{lm}(\theta'\varphi') Y_{lm}(\theta\varphi) \qquad (5.6)$$

And we write

$$\exp[qq'+i\vec{r}.\vec{r}'] = 4\pi \sum_{l=0}^{\infty} \sum_{m=-l}^{+l} i^l [\exp[vq'\cos\chi]$$
$$\left(\frac{\pi}{2kr}\right)^{1/2} J_{l+1/2}(r'v\sin\chi) Y^*_{lm}(\theta'\varphi') Y_{lm}(\theta\varphi)] \qquad (5.7)$$

With the second generating function of Gegenbauer polynomials

$$e^{z\cos\chi}(\frac{z}{2}\sin\chi)^{\frac{1}{2}-\alpha} J_{\alpha-\frac{1}{2}}(z\sin\chi) = \sum_{n=0}^{\infty} \frac{\Gamma(2\alpha)}{\Gamma(\alpha+\frac{1}{2})\Gamma(2\alpha+n)} C_n^{(\alpha)}(\cos\chi) z^n \qquad (5.8)$$

We obtain

$$\exp[qq'+i\vec{r}.\vec{r}'] = 4\pi \sum_{n=0}^{\infty}\sum_{n=0}^{\infty} \frac{(2l+1)\Gamma(2\alpha)}{\Gamma(\alpha+\frac{1}{2})\Gamma(2\alpha+n)} i^l \sum_{l=0}^{\infty} [C_n^{(\alpha)}(\cos\chi)(r'v)^n \times$$
$$\sum_{m=-l}^{+l} Y^*_{lm}(\theta'\varphi') Y_{lm}(\theta\varphi)] \qquad (5.9)$$

With Legendre duplication formula

$$\Gamma(\frac{1}{2})\Gamma(2n+2) = 2^{2n}\Gamma(n+\frac{3}{2})\Gamma(n+1) \qquad (5.10)$$

and we multiply (5.9) and (5.5) by $\int Y_{lm}(\theta'\varphi') d\Omega'$ and then we execute the integration.

We obtain

$$\frac{\pi^{3/2}}{2^{l-1/2}} (2l+1) i^l \frac{\Gamma(2(l+1))}{\Gamma(l+3/2)\Gamma(2l+2+n)} C_{n-l-1}^{l+1}(\cos\chi) Y_{lm}(\theta\varphi)$$
$$= \sum_{\mu\mu'} \int [\varphi_{j\mu_1}(z')\varphi_{j\mu_2}(\bar{z}')] Y_{lm}(\theta'\varphi') d\Omega'_k ] D^j_{(m,m')}(U_2) \qquad (5.11)$$

With the help of the expression [17]

$$\frac{1}{8\pi^2} \int dU_2 D^{\frac{n-1}{2}}_{(\frac{n-1}{2},m_1)}(U_2) D^{\frac{n-1}{2}}_{(-\frac{n-1}{2},m_2)}(U_2) D^l_{(0,m)}(U_2) =$$
$$\begin{pmatrix} (n-1)/2 & (n-1)/2 & l \\ (n-1)/2 & -(n-1)/2 & 0 \end{pmatrix} \begin{pmatrix} (n-1)/2 & (n-1)/2 & l \\ m_1 & m_2 & m \end{pmatrix} \qquad (5.12)$$

Finally we get the well known [2] expression

$$Y_{nlm}(\vec{v}) = \frac{(-i)^l}{\pi}(\frac{n}{2})^{\frac{1}{2}}\sum_{m_1 m_2}(-1)^{\frac{(n-1)}{2}-m_2}(2l+1)^{\frac{1}{2}}\begin{pmatrix}\frac{n-1}{2} & \frac{n-1}{2} & l \\ m_1 & -m_2 & m\end{pmatrix}D^{\frac{n-1}{2}}_{(m_1,m_2)}(U_2) \quad (5.13)$$

### 5.2 Integral representation of the generating functions of Gegenbauer polynomials.

With the development of the plane wave and the expression (5.6) we find that

$$\int \exp[\beta(\vec{a}.\vec{r}')/2 + \alpha(qq'+i\vec{r}\cdot\vec{r}')]d\mu(u') = (1/2\pi)\sum_{l=0}^{\infty}\sum_{m=-l}^{+l}i^l\{\int[\exp[\alpha vq'\cos\chi]\times$$

$$j_l(vq'\alpha\sin\chi)(\alpha\beta q')^l q' dq'\}\frac{(\vec{a}\cdot\vec{r})^l}{2^l l!}] \quad (5.14)$$

We can execute the integration using the Bargmann integral and we repeat the same calculation of paragraph 4. We obtain

$$\int \exp[\beta(\vec{a}.\vec{r}')/2 + \alpha(qq'+i\vec{r}\cdot\vec{r}')]d\mu(u') = \int \exp[-\bar{z}^{\prime t} X z']$$

With $X = \begin{pmatrix} 1-(\alpha q + i\alpha z + \frac{\beta}{2}a_z) & -(i\alpha x + \frac{\beta}{2}a_x + \alpha y - i\frac{\beta}{2}a_y) \\ -(i\alpha x + \frac{\beta}{2}a_x - \alpha y + i\frac{\beta}{2}a_y) & 1-(\alpha q + i\alpha z + \frac{\beta}{2}a_z) \end{pmatrix}$

And $\det(X) = 1 - 2\alpha q + \alpha^2 \vec{v}^2 - i\alpha\beta(\vec{a}\cdot\vec{r})$

Then $\int \exp[\beta(\vec{a}.\vec{r}')/2 + \alpha(qq'+i\vec{r}\cdot\vec{r}')]d\mu(u') = \frac{1}{1-2\alpha q + \alpha^2 \vec{v}^2 - i\alpha\beta(\vec{a}\cdot\vec{r})}$

$$= \sum_l \frac{(-i\alpha\beta)^l(\vec{a}\cdot\vec{r})^l}{(1-2\alpha q + \alpha^2\vec{v}^2)^{l+1}} \quad (5.15)$$

Put $v=1, t=u'^2$, and comparing the two expressions (5.14) and (5.15) we
Find the integral representation of the generating functions of Gegenbauer:

$$\frac{1}{(1-2\alpha\cos\chi+\alpha^2)^{l+1}} = \frac{(-1)^l}{\pi 2^{l+1}l!}\int_0^\infty \exp[t\cos\chi]j_l(t\sin\chi)t^{l+1}e^{-t}dt \quad (5.16)$$

## 6. Clifford algebra and generating function of Gegenbauer polynomials

We noticed a relationship between the generating function of Gegenbauer polynomials and the octonions algebra and this part aims to present this relationship.

### 6.1 Bargmann integral and Levi-Civita transformation

The quadratic transformation $R^2 \to R^2$ is

$$x' = 2u_1 u_2, \; y' = u_1^2 - u_2^2, \; r' = u_1^2 + u_2^2 \quad (6.1)$$

put $(x,y,z) = (x_1, x_2, x_3)$
And by analogy with the expression (5.3) we write:

$$zr' + ixx' + iyy' = \begin{pmatrix} u_1 & u_2 \end{pmatrix} A_1 \begin{pmatrix} u_1 \\ u_2 \end{pmatrix} = \begin{pmatrix} u_1 & u_2 \end{pmatrix} \begin{pmatrix} (x_3 + ix_2) & ix_1 \\ ix_1 & (x_3 - ix_2) \end{pmatrix} \begin{pmatrix} u_1 \\ u_2 \end{pmatrix} \quad (6.2)$$

The Bargmann integral in this case is

$$\int \exp[x_3 r' + ix_1 x' + ix_2 y'] d\mu(u) = \frac{1}{\sqrt{(1 - 2x_3 \alpha + \alpha^2 r^2)}}, \quad (6.3)$$

$$d\mu(u) = e^{-\Sigma u_i^2} \prod du_i, \quad r^2 = x_3^2 + x_2^2 + x_1^2$$

The second part is the generating function of Gegenbauer polynomials

We also write $A_1 = \begin{pmatrix} (x_3 + ix_2) & ix_1 \\ ix_1 & (x_3 - ix_2) \end{pmatrix} = x_3 I + x_2 \Gamma_2 + x_1 \Gamma_1$ (6.4)

With $\quad \Gamma_2^2 = \Gamma_1^2 = -I$

**6.2 Bargmann integral and the quaternions**

It is well known that $A_2 = \begin{pmatrix} x_4 + ix_3 & x_2 + ix_1 \\ -x_2 + ix_1 & x_4 - ix_3 \end{pmatrix} = \sum_{i=1}^{4} x_i \Gamma_i, \quad \Gamma_4 = I$ (6.5)

Where $(I, \Gamma_3, \Gamma_2, \Gamma_1)$ are the representations matrices of quaternion. We find by a direct calculation the Bargmann integral

$$\int \exp\left[\alpha \begin{pmatrix} \bar{z}_1 & \bar{z}_2 \end{pmatrix} \begin{pmatrix} x_4 + ix_3 & x_2 + ix_1 \\ -x_2 + ix_1 & x_4 - ix_3 \end{pmatrix} \begin{pmatrix} z_1 \\ z_2 \end{pmatrix} \right] d\mu(z) = \frac{1}{(1 - 2x_4 \alpha + \alpha^2 x_4^2)}, \quad (6.6)$$

$$d\mu(z) = e^{-\Sigma u_i^2} \prod du_i$$

The second part is the generating function of Gegenbauer polynomials

**6.3 Bargmann integral and the Octonions**

The Hurwitz transformation of $R^8 \to R^5$ is given by

$$(x_1 + ix_2) = 2(\bar{z}_1 z_3 + z_2 \bar{z}_4), \quad (x_3 + ix_4) = 2(\bar{z}_1 z_4 - z_2 \bar{z}_3)$$

$$r_1 = u_1^2 + u_2^2 + u_3^2 + u_4^2, \quad r_2 = u_5^2 + u_6^2 + u_7^2 + u_8^2 \quad (6.7)$$

$$x_5 = r_1 - r_2, \quad r = r_1 + r_2$$

We consider as previously:

$$(\bar{z}^t A_3 z) = x_6 r + i(x_1 x_1' + x_2 x_2' + x_3 x_3' + x_4 x_4' + x_5 x_5') =$$

$$(\bar{z})^t A_3(z) = (\bar{z})^t \begin{pmatrix} x_6 + ix_5 & 0 & -x_1 + ix_2 & -x_4 + ix_3 \\ 0 & x_6 + ix_5 & -x_4 - ix_3 & x_1 + ix_2 \\ x_1 + ix_2 & x_4 - ix_3 & x_6 - ix_5 & 0 \\ x_4 + ix_3 & -x_1 + ix_2 & 0 & x_6 - ix_5 \end{pmatrix} (z) \quad (6.8)$$

t is the transpose and $\quad (z)^t = (z_1 \quad z_2 \quad z_3 \quad z_4)$

The Bargmann integral in this case is

$$\int \exp(\alpha \bar{z}^t A_3 z) d\mu(z) = \frac{1}{(1 - 2\alpha x_6 - \alpha^2 u^2)^2} \quad (6.9)$$

we find again that the second member is the generating function of Gegenbauer polynomials.

We also write
$$A_3 = \sum_i x_i \Gamma_i, \quad \Gamma_i^2 = -1$$
$$\Gamma_i \Gamma_j + \Gamma_j \Gamma_i = \delta_{ij}$$
(6.10)

We note that $\{\Delta_6 \exp[\alpha z^t A_3 z] = 0\}$ which permit us to do the expansion of (6.8) on the basis of SO (6) or on the basis of SU (3) and the matrix of passage between these bases is determined by the same method of part 5.

**6.4 Bargmann integral and the Clifford Algebra**

Based on the expression of $A_1$, $A_2$ and $A_3$ we can generalize these results by writing:
$$A_n = \begin{pmatrix} (x_{2n} + ix_{2n-1})I_{2^{n-2}} & A_{n-1} \\ -\bar{A}_{n-1}^t & (x_{2n} - ix_{2n-1})I_{2^{n-2}} \end{pmatrix}$$
(6.11)

Using a symbolic program we find also:

For n=4  $\int \exp(\alpha \bar{z}^t A_4 z) d\mu(z) = \dfrac{1}{(1 - 2\alpha x_8 - \alpha^2 u^2)^4}$  (6.12)

For n=5  $\int \exp(\alpha \bar{z}^t A_5 z) d\mu(z) = \dfrac{1}{(1 - 2\alpha x_{10} - \alpha^2 u^2)^8}$  (6.13)

And
$$A_n = \sum x_i \Gamma_i,$$
$$\Gamma_i^2 = -1, \Gamma_i \Gamma_j + \Gamma_j \Gamma_i = \delta_{ij}$$
6.14)

We deduce from the above mentioned that there is a close relationship between the Clifford algebra and the generating functions of Gegenbauer polynomials and our method of calculating Fourier transformation of the position coordinates can be generalized to any orders.